\documentclass[preprint,superscriptaddress,amsmath,amssymb]{revtex4-1}
\usepackage{dcolumn}
\usepackage{graphicx}
\usepackage{float}
\usepackage{physics}
\usepackage{xcolor}
\usepackage[colorlinks=true,allcolors=blue]{hyperref}

\usepackage{amsmath}
\usepackage{color}
\usepackage{amssymb}
\usepackage{bm}

\allowdisplaybreaks[1]
\usepackage{bm}
\usepackage[]{changes}
\usepackage{extarrows}
\renewcommand\sout{\bgroup \color{red} \ULdepth=-.5ex \ULset}

\graphicspath{{figs/}}
\begin{document}

\title{Vector meson dominance in photon structure functions at small $x$ from holography}

\author{Wei Gao}\email{weigao@my.swjtu.edu.cn}
\affiliation{School of Physical Science and Technology, Southwest Jiaotong University, Chengdu, 610031, People's Republic of China}

\author{Siming Liu}
\email{liusm@swjtu.edu.cn}
\affiliation{School of Physical Science and Technology, Southwest Jiaotong University, Chengdu, 610031, People's Republic of China}

\author{Wenbin Lin}
\email{lwb@usc.edu.cn}
\affiliation{School of Physical Science and Technology, Southwest Jiaotong University, Chengdu, 610031, People's Republic of China}
\affiliation{School of Mathematics and Physics, University of South China, Hengyang 421001, People's Republic of China}

\author{Akira Watanabe}
\email{watanabe.akira@oshima-k.ac.jp}
\affiliation{National Institute of Technology, Oshima College, Oshima 742-2193, Japan}

\date{\today}

\begin{abstract}
We investigate the photon structure functions via the photon-photon and photon-vector meson scattering within the framework of holographic QCD, focusing on the small Bjorken $x$ region and assuming that the Pomeron exchange dominates. The quasi-real photon structure functions are formulated as the convolution of the known $\mathrm{U(1)}$ vector field wave function with the Brower-Polchinski-Strassler-Tan (BPST) Pomeron exchange kernel in the five-dimensional AdS space.
Assuming the vector meson dominance, the photon structure functions can be calculated in a different way with the BPST kernel and vector meson gravitational form factor, which can be obtained in a bottom-up AdS/QCD model, for the Pomeron-vector meson coupling.
It is shown that the obtained $F_2$ structure functions in the both ways agree with the experimental data, which implies the realization of the vector meson dominance within the present model setup.
Calculations for the longitudinal structure function and the longitudinal-to-transverse ratio are also presented.
\end{abstract}

\maketitle
    
\section{Introduction}
As an elementary particle, the photon fundamentally differs from non-perturbative composite particles such as hadrons. However, in a hard scattering process, an energetic photon can fluctuate into quark-antiquark pairs, which further radiate gluons. Over the past decade, studying photon structure has become a meaningful way to test the effectiveness of perturbative quantum chromodynamics (QCD). The feasibility of electron-photon deep inelastic scattering (DIS) was first explored and its cross section estimated in Refs.~\cite{Walsh:1971xy,Brodsky:1971vm}. In particular, experimental measurements made at the Large Electron-Positron Collider (LEP) have provided valuable data for the photon structure function~\cite{OPAL:2000nfx,OPAL:2002vci}. In $1977$, Witten first evaluated the leading-order QCD corrections~\cite{Witten:1977ju}. Subsequent higher-order calculations based on perturbative QCD were developed~\cite{LlewellynSmith:1978syc,Bardeen:1978hg,Duke:1980ij,Antoniadis:1982fv,Gluck:1983mm,Fontannaz:1992gj,Gluck:1991ee}, and next-to-next-to-leading-order results were obtained~\cite{Moch:2001im}. The consistency between theoretical results and experimental data effectively verifies the reliability of perturbative QCD calculations. When the Bjorken scaling variable $x$ is less than $0.1$, a photon can fluctuate into vector mesons, so the hadronic contributions become non-negligible in electron-photon DIS. In particular, when $x$ is less than $0.01$, the hadronic component dominates, and the photon is no longer treated as a point particle but instead behaves like a vector meson. This behavior can be described by an effective model, the so-called vector meson dominance model proposed by Sakurai in 1960~\cite{Sakurai:1960ju}.

Holographic QCD based on the anti-de Sitter/conformal field theory (AdS/CFT) correspondence~\cite{Taliotis:2009ne,Gubser:1998bc,Witten:1998qj} is an effective model. Through the AdS/CFT correspondence, strongly coupled gauge theory processes can be mapped to weakly coupled gravitational theory in the higher dimensional AdS space. Since the pioneering introduction of the AdS/CFT correspondence by Maldacena in 1997, its applications to QCD processes have been attempted intensively~\cite{Kruczenski:2003be,Kruczenski:2003uq,Son:2003et,Sakai:2004cn,Sakai:2005yt,Erlich:2005qh,DaRold:2005mxj,Erdmenger:2007cm}.
The theory introduces the QCD scale $\Lambda_{\mathrm{QCD}}$ and breaks conformal symmetry in the strong coupling limit. In the study of hadron physics, the model has achieved remarkable success in phenomenological applications, such as mass spectra~\cite{Erlich:2005qh,Karch:2006pv,Gherghetta:2009ac}, form factors~\cite{Brodsky:2006uqa,Grigoryan:2007my,Abidin:2009hr}, and high-energy scattering~\cite{Polchinski:2001tt,Brower:2006ea}.

In this paper, we investigate the structure functions of the photon at small $x$ via the photon-photon ($\gamma\gamma$) and photon-$\rho$ meson ($\gamma\rho$) scattering, taking into account the Pomeron exchange in holographic QCD. The Pomeron is considered as a color-singlet gluonic state formed by the multi-gluon exchange.
It is shown that total cross sections of various hadron-hadron scattering processes can be described by the Pomeron exchange in Ref.~\cite{Donnachie:1992ny}. In holographic QCD, the Pomeron corresponds to the Reggeized graviton in the five-dimensional AdS space. Polchinski and Strassler studied high-energy scattering based on the AdS/CFT correspondence~\cite{Polchinski:2001tt}, and subsequently, many related studies have been done~\cite{Polchinski:2002jw,Boschi-Filho:2005xct,Hatta:2007he,BallonBayona:2007rs,Pire:2008zf,Stoffers:2012zw}. In Ref.~\cite{Brower:2006ea}, the authors evaluated the graviton exchange contribution, and proposed the BPST Pomeron exchange kernel. Extensive studies in Refs.~\cite{Cornalba:2007zb,Cornalba:2008sp,Brower:2007qh,Brower:2007xg,Cornalba:2009ax,Cornalba:2010vk,Cornalba:2010vk} demonstrated that this kernel works for the analysis on the DIS structure functions in the small $x$ region, and various phenomenological studies have been done successfully~\cite{Brower:2010wf,Watanabe:2012uc,Agozzino:2013zgy,Watanabe:2013spa,Watanabe:2015mia,Watanabe:2018owy,Watanabe:2019zny}.

Following a similar approach, the photon structure functions at small $x$ are expressed as the convolution of the BPST kernel with the wave functions of the incident photon and target photon in the five-dimensional AdS space. The photon in the space can be regarded as the five-dimensional $\mathrm{U(1)}$ vector field coupled to the lepton at the ultraviolet (UV) boundary. The effectiveness of the BPST kernel is verified by comparison with experimental data. Our proposed photon structure functions involve only one free parameter, significantly reducing the model dependence, which makes the theoretical results more concise and increases the predictive power. The results of the photon structure function $F_2(x, Q^2)$ are consistent with the LEP data. In addition, the obtained results are consistent with the photon parton distribution functions (PDFs) proposed by Gl$\ddot{\mathrm{u}}$ck, Reya, and Schienbein~\cite{Gluck:1999ub}. Since they include hadronic components that dominate at small $x$, this consistency may suggest that the vector meson dominance is realized in the current model setup.

This paper is organized as follows. In Sec.~\ref{2}, we show the whole setup of the model by introducing the BPST Pomeron exchange kernel and give the different overlap functions in the cases of the $\gamma\gamma$ and $\gamma\rho$ scattering. In Sec.~\ref{3}, we give numerical results for the structure functions. Finally, the summary is given in Sec.~\ref{4}.

\section{Model setup}\label{2}
The differential cross section of the unpolarized electron-photon DIS is expressed with the two structure functions, $F_2^{\gamma}(x,Q^2)$ and $F_L^{\gamma}(x,Q^2)$, as
\begin{eqnarray}
&& \frac{d^2\sigma_{e\gamma \to eX}}{dxdQ^2} = \frac{2\pi \alpha^2}{xQ^4}\left\{\left[1+(1-y)^2\right]F_2^{\gamma}(x,Q^2)-y^2F_L^{\gamma}(x,Q^2)\right\}~,\label{sigam}
\end{eqnarray}
where $\alpha$ represents the fine structure constant and $y$ denotes the inelasticity.
The Bjorken scaling variable $x$ is defined as
\begin{eqnarray}
&& x=\frac{Q^2}{Q^2+W^2+P^2}~,\label{x}
\end{eqnarray}
where $W$ denotes the invariant mass of the final hadronic state, $Q^2=-q^2$, and $P^2=p^2$, in which $q$ and $p$ are the four-momenta of the probe photon and the target photon, respectively.
We focus on the small $x$ regime, in which the condition $W^2\gg Q^2 \gg P^2$ is satisfied and then the above expression can be approximated as $x\approx Q^2/W^2$.
The vector meson dominance model~\cite{Sakurai:1960ju}, applicable in this region, indicates that the target photon may exhibit vector meson-like behaviour rather than that of a point particle.

The photon structure functions in Eq.~(\ref{sigam}) are expressed in the five-dimensional AdS space as~\cite{Brower:2007qh,Brower:2007xg}
\begin{eqnarray}
&& F_i^{\gamma}(x,Q^2) = \frac{\alpha g_0^2\rho^{3/2}Q^2}{32\pi^{5/2}}\int dzdz'P_{13}^{(i)}(z,Q^2)P_{24}(z',P^2)(zz')\mathrm{Im}\left[\chi_{c}(W^2,z,z')\right]~,\label{F_i}
\end{eqnarray}
with $i=2,L$, where the parameters $g_0^2$ and $\rho$ are adjustable and they govern the magnitude and energy dependence of the structure functions, respectively. The overlap functions $P_{13}^{(i)}(z,Q^2)$ and $P_{24}(z',P^2)$ describe the density distributions of the incident and target particles in the AdS space, respectively. In the preceding studies~\cite{Brower:2010wf,Watanabe:2012uc}, it was demonstrated that incorporating the confinement effect in QCD is essential for reproducing the proton structure function data with the BPST kernel, except in the high $Q^2$ region. Consequently, to numerically evaluate the structure functions, we employ the modified kernel rather than the conformal kernel.
\begin{eqnarray}
&& \mathrm{Im}\left[\chi_{mod}(W^2,z,z')\right] = \mathrm{Im}\left[\chi_{c}(W^2,z,z')\right]+\mathcal{F}(z,z',\tau)\mathrm{Im}\left[\chi_{c}(W^2,z,z_0^2/z')\right]~,\label{chi-mod}
\end{eqnarray}
\begin{eqnarray}
&& \mathcal{F}(z,z',\tau) = 1-2\sqrt{\rho \tau }e^{\eta^2}\mathrm{erfc}(\eta)~,\label{F}
\end{eqnarray}
\begin{eqnarray}
&& \eta = \left(-\log{\frac{zz'}{z_0^2}}+\rho\tau\right)/\sqrt{\rho\tau}~.\label{eta}
\end{eqnarray}
The first term on the right-hand side of Eq.~(\ref{chi-mod}) corresponds to the conformal kernel, which is given by~\cite{Brower:2007xg,Brower:2006ea}
\begin{eqnarray}
&& \mathrm{Im}\left[\chi_{c}(W^2,z,z')\right] = e^{(1-\rho)\tau}e^{-\left[\left(\log z-\log z'\right)^2/\rho\tau\right]}/{\sqrt{\tau}} ~,\label{chi-c}
\end{eqnarray}
where $\tau = \log(\rho zz'W^2)/2$. The second term on the right-hand side of Eq.~(\ref{chi-mod}) mimics the confinement effect adjusted by the parameter $z_0$.

The overlap functions $P_{13}(z,Q^2)$ and $P_{24}(z',P^2)$, which represent the density distributions of the incident and target particles within the AdS space, respectively, satisfy the normalization condition $\int dz P_{ij}(z) = 1$ in the on-shell case. In the present study, they depend on the virtualities $Q^2$ and $P^2$ for off-shell photons. $P_{13}^{(L)}(z, Q^2)$ denotes the wave function of the longitudinally polarized photon, while $P_{24}(z',P^2)$ adopts the same functional form as $P_{13}(z, Q^2)$. The massless five-dimensional $\mathrm{U(1)}$ vector field can be identified as the physical photon at the UV boundary, satisfying the Maxwell equation in the five-dimensional AdS background spacetime. We apply the wave functions with a weight $w$ on its longitudinal component~\cite{Watanabe:2019zny}
\begin{eqnarray}
&& P_{13}^{(2)}(z,Q^2) = zQ^2\left[wK_0^2(Qz)+K_1^2(Qz)\right]~,\label{P-13-2}\\
&& P_{13}^{(L)}(z,Q^2) = wzQ^2K_0^2(Qz)~,\label{P-13-L}
\end{eqnarray}
where $w=0.6198$, $K_0(Qz)$ and $K_1(Qz)$ are the modified Bessel functions of the second kind. The $w = 1$ result is presented in Ref.~\cite{Polchinski:2002jw} and satisfies the Maxwell equation in the bulk AdS space, providing a theoretical foundation for analyzing proton structure functions. The known results help reproduce the experimental data of proton $F_2$; however, the obtained $F_L$ is larger than the data. Considering the longitudinal-to-transverse ratio of $R_{L/T} = F_L/F_T$, the experimental result is $R_{L/T} = 0.26$~\cite{H1:2010fzx}. Therefore, in this study we apply the weight $w$ to $P_{13}^{(2)}(z,Q^2)$ and $P_{13}^{L}(z,Q^2)$. This factor is determined with experimental data to refine the theoretical predictions for $F_L$. The goal is to improve agreement between theory and experiment by enhancing the longitudinal structure function's accuracy while preserving the transverse component's validity.

Regarding the vector meson dominance model, we examine photon-$\rho$ meson scattering, where $P_{24}^{\rho}(z')$ is described by the $\rho$ meson gravitational form factor in Ref. \cite{Abidin:2008ku}.
\begin{eqnarray}
&& P_{24}^{\rho}(z') = \frac{1}{z'}\psi_n(z')\psi_n(z')~,\label{P-24-prho}
\end{eqnarray}
where
\begin{eqnarray}
&& \psi_n(z') = \frac{\sqrt{2}}{z_0J_1(m_nz_0)}z'J_1(m_nz')~,\label{Psi-n}
\end{eqnarray}
in which $z_0=1/{\Lambda_{\mathrm{QCD}}} = 1/322$ MeV$^{-1}$ and $m_n = m_{\rho} = 775.36$ MeV.
\begin{figure}[!t]
\centering
\begin{tabular}{ccccc}
\includegraphics[scale=0.65]{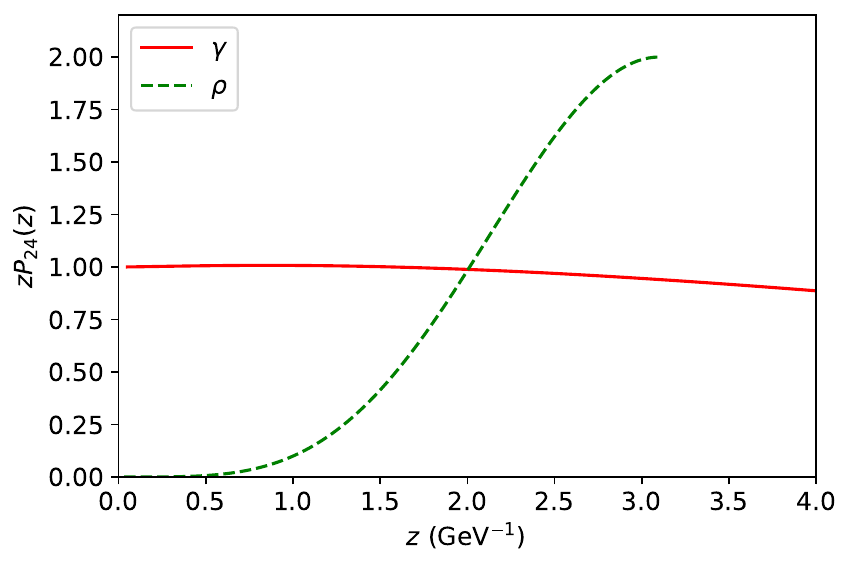}
\end{tabular}
\caption{The overlap functions $zP_{24}(z)$ in the integrand of Eq.~(\ref{F_i}). The photon is shown by the red solid curve, and one for the $\rho$ meson by the green dashed curve.}
\label{zP24}
\end{figure}
In Fig.~\ref{zP24}, the overlap functions $zP_{24}(z)$ for the photon, shown by the red solid line, and the $\rho$-meson, represented by the green dashed line, are displayed. The photon exhibits UV-dominated behavior $zP_{24}^\gamma(z) \approx 1.0$, while the $\rho$-meson shows IR-localized enhancement due to the non-perturbative confinement.

\section{Numerical results}\label{3}
In this study, we adopt the BPST kernel parameter as outlined in Ref.~\cite{Watanabe:2019zny}, with $z_0 = 6$ $\text{GeV}^{-1}$ for the modified kernel. This method has been shown to accurately represent the nucleon structure function data at small values of $x$. Since the coefficient $g_0^2$ in Eq.~(\ref{F_i}) is unknown, calibration is required to align with the absolute magnitude of the data. There are limited experimental data on the real photon structure function at small $x$. In the range $ x \leq 0.025 $, we considered nine data points from the OPAL Collaboration at LEP~\cite{OPAL:2000nfx}, and by performing data fitting using the MINUIT package~\cite{James:1975dr}, we determined $ g_0^2 = 23.82 \pm 1.49 $ and $ 72.58 \pm 4.51$ for the $\gamma\gamma$ and $\gamma\rho$ cases, respectively. The virtuality $P^2$ of the target photon should be quite small, but it cannot be precisely zero. The measurement range tells $P^2 \sim \mathcal{O}(0.01)~\text{GeV}^2 $~\cite{Nisius:1999cv,Krawczyk:2000mf}, so we assume $ P^2 = 0.01~\text{GeV}^2 $~\cite{Gluck:1999ub}.

We here give an explanation for the treatment of the conformal invariant $\tau$, which appears in the kernels of Eqs.~(\ref{chi-mod}) and (\ref{chi-c}). Through straightforward integration, we isolate the contribution from the regions corresponding to small values of $z$ and $z'$, where the photon distributions are most concentrated. This process results in a negative value for $\tau$, indicating the presence of an imaginary component within the kernels.
This phenomenon is inherent to the BPST Pomeron kernel, as discussed in Refs.~\cite{Brower:2007xg,Brower:2006ea}. This issue was noted in earlier works like Ref.~\cite{Brower:2010wf} and handled via schemes~\cite{Watanabe:2015mia}.
This imaginary part is a direct consequence of the behaviour of the photon distributions in these regions and plays a crucial role in shaping the overall structure of the kernels. Therefore, in the definition of $\tau$, the variables $z$ and $z'$ are assigned their respective average values, which are defined as
\begin{eqnarray}
&& \Bar{z} = \frac{\int z^2P_{13}^{(i)}(z,Q^2)dz}{\int zP_{13}^{(i)}(z,Q^2)dz},~~~~~~\Bar{z}' = \frac{\int z'^2P_{24}(z',Q^2)dz'}{\int z'P_{24}(z',Q^2)dz'},\label{zza}
\end{eqnarray}
respectively, and considering the product $zz'$ as the characteristic scale $\Bar{z}\Bar{z}^{'}$ in the kernels.

\begin{figure}[!t]
\centering
\begin{tabular}{ccccc}
\includegraphics[scale=0.65]{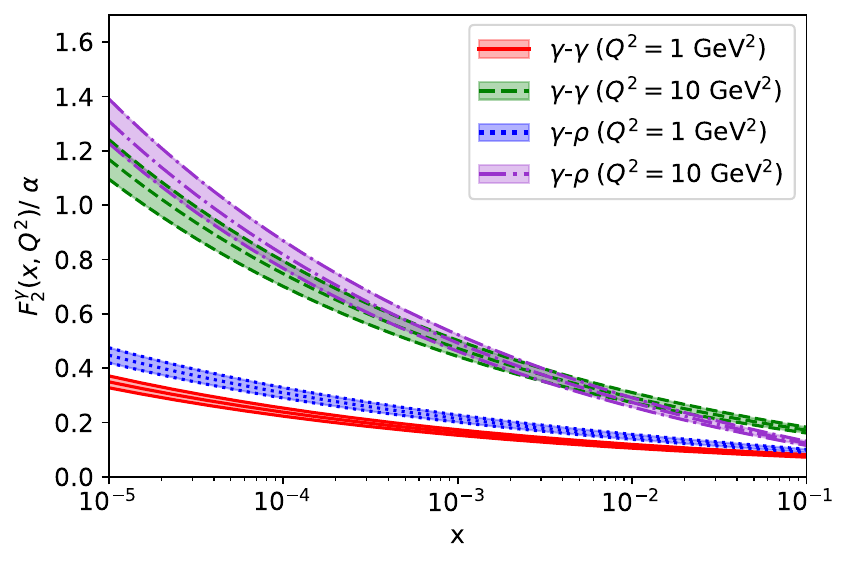}
\end{tabular}
\caption{The structure function $F_2^\gamma(x, Q^2)$ as a function of $x$ for $Q^2 = 1$ and $10$~$\text{GeV}^2$. The parameter value $z_0 = 6.0$~$\text{GeV}^{-1}$ is used to generate the curves with the modified BPST kernel. The uncertainties for $g_0^2$ are shown in the figure.}
\label{F2alpha}
\end{figure}

Figure~\ref{F2alpha} shows the variation of the structure function $F_2^{\gamma}(x, Q^2)$ as a function of the Bjorken variable $x$ for the $\gamma\gamma$ and $\gamma\rho$ cases.
All the four results decrease with $x$ and the larger $Q^2$ results are above the lower $Q^2$ results, which are consistent with the experimental facts.
Focusing on the results with $Q^2 = 10$~GeV$^2$, it is found that the $\gamma \rho$ result is slightly larger than the $\gamma \gamma$ result in the very small $x$ region, but this order becomes opposite in the relatively larger $x$ region.

\begin{figure}[!t]
\centering
\begin{tabular}{ccccc}
\includegraphics[scale=0.60]{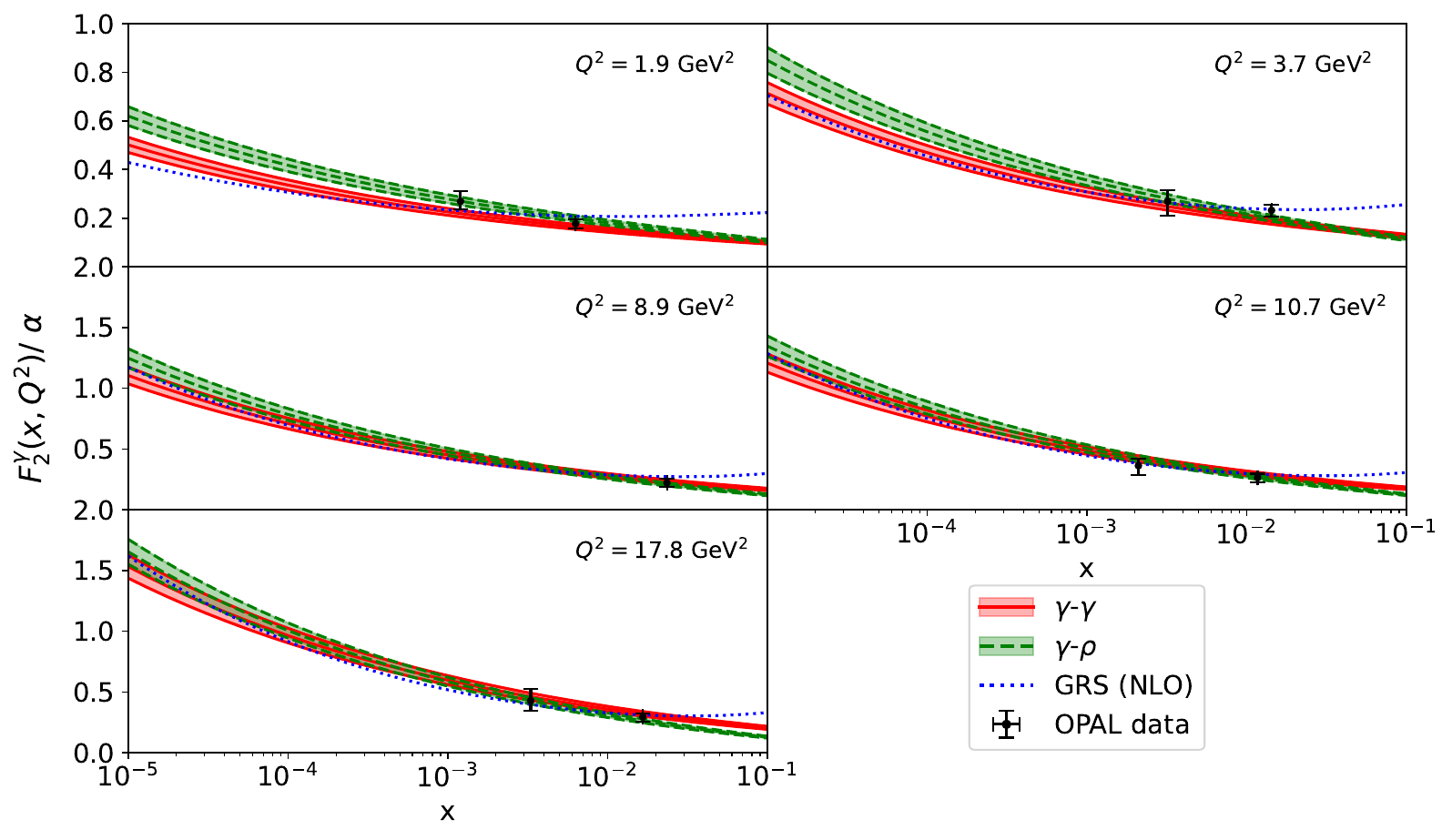}
\end{tabular}
\caption{The structure function $F_2^\gamma(x, Q^2)$ as a function of $x$ for various values of $Q^2$, compared with the experimental data from OPAL Collaboration~\cite{OPAL:2000nfx} and those calculated from the GRS PDF set~\cite{Gluck:1999ub}. In each panel, the red solid and green dashed curves represent our calculations, while the blue dotted curves indicate the GRS predictions. The uncertainties for $g_0^2$ are shown in the figure.}
\label{DISwz2}
\end{figure}

In Fig.~\ref{DISwz2}, we present the resulting structure function $F^{\gamma}_2(x, Q^2)$ for various $Q^2$ values (1.9, 3.7, 8.9, 10.7, and 17.8~$\text{GeV}^2$), compared with the experimental data from OPAL Collaboration and those calculated from the well-known GRS PDF set~\cite{Gluck:1999ub}.
The solid red and green dashed curves represent our calculations for the two scattering cases, while the dotted blue curves represent the GRS (NLO) predictions, serving as a reference model. Our observations indicate that as $x$ increases, the value of $F_2^\gamma(x, Q^2)$ generally decreases, consistent with the expected behavior of PDFs at small $x$.
Although the number of the available data points is quite limited, our both results are in agreement with them.
It is also found that our results are consistent with the GRS predictions in the considered kinematic regions.
For our results at $Q^2 = 1.9$ and $3.7$~GeV$^2$, deviations between the $\gamma \gamma$ and $\gamma \rho$ cases are seen.
To pin down this, more experimental date are necessary.

\begin{figure}[!t]
\centering
\begin{tabular}{ccccc}
\includegraphics[scale=0.65]{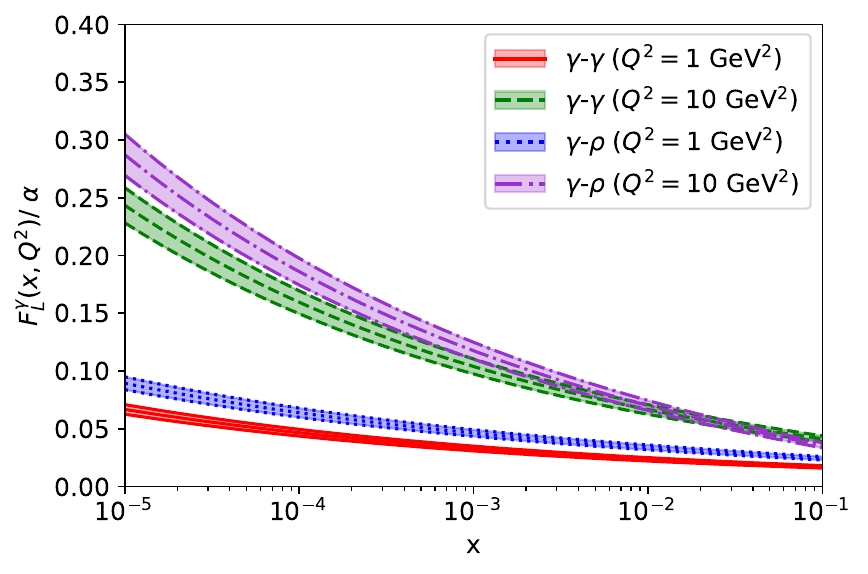}
\end{tabular}
\caption{The longitudinal structure function $F_L^\gamma(x, Q^2)$ as a function of $x$ for $Q^2 = 1$ and $10$~$\text{GeV}^2$. The parameter value $z_0 = 6.0$~$\text{GeV}^{-1}$ is used to obtain the curves with the modified kernel. The uncertainties for $g_0^2$ are shown in the figure.}
\label{F_Lalpha}
\end{figure}

Figure~\ref{F_Lalpha} shows the resulting longitudinal structure function $F_L^\gamma(x, Q^2)$ for the $\gamma\gamma$ and $\gamma\rho$ cases at $Q^2 = 1$ and $10$~$\text{GeV}^2$.
Overall, the observed behaviours are similar to those for the $F_2$ structure function.
For the both energy scales, the $\gamma\rho$ results are larger than the $\gamma\gamma$ results.
\begin{figure}[!tb]
\centering
\begin{tabular}{ccccc}
\includegraphics[scale=0.65]{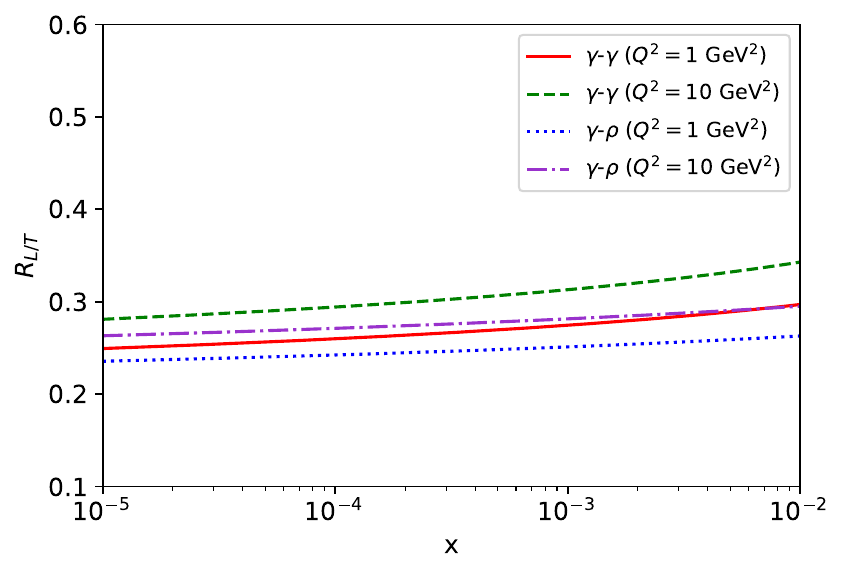}
\end{tabular}
\caption{The longitudinal-to-transverse ratio $ R_{L/T} = F_{\gamma}^L(x, Q^2)/F_{\gamma}^T(x, Q^2) $ as a function of $x$ for $Q^2 = 1$ and $10$~$\text{GeV}^2$. The parameter value $z_0 = 6.0$~$\text{GeV}^{-1}$ is used to obtain the curves with the modified kernel.}
\label{RLT}
\end{figure}
Finally, we present in Fig.~\ref{RLT} the $x$ dependence of the longitudinal-to-transverse ratio for the structure functions, defined as $R_{L/T} = {F^\gamma_L(x, Q^2)}/{F^\gamma_T(x, Q^2)}$, where the transverse structure function is given by $F_T^\gamma = F_2^\gamma - F_L^\gamma$. The results are presented for two distinct $Q^2$ values, $Q^2 = 1 $ and $ 10~\text{GeV}^2$, considering the two cases, $\gamma\gamma$ and $\gamma\rho$.  From the figure, it is evident that $R_{L/T}$ increases with both $x$ and $Q^2$.
Similar to our results of the $F_2$ and $F_L$ structure functions, deviations between the $\gamma\gamma$ and $\gamma\rho$ results at the fixed value of $Q^2$ can be seen.
To further constrain the present descriptions, more experimental data, especially at smaller $x$, are needed.

\section{Summary}\label{4}
We have investigated the real photon structure functions $F_2$ and $F_L$ in the small Bjorken $x$ region via the $\gamma\gamma$ and $\gamma\rho$ scattering within the framework of holographic QCD.
For the $\gamma\gamma$ case, the structure functions are calculated by combining the BPST Pomeron exchange kernel with the wave functions of the $\mathrm{U(1)}$ vector field in the five-dimensional AdS space.
For the $\gamma\rho$ case, the $\rho$ meson gravitational form factor, which can be obtained with a bottom-up AdS/QCD model, is used for the Pomeron coupling with the target particle.
In earlier studies on the nucleon DIS, two of the three adjustable parameters in the Pomeron exchange kernel were fixed with basic hadron properties such as the hadron mass, leaving the third one as a free parameter, which is finely tuned with the experimental data of the structure functions in this study.

We have demonstrated that our calculations of the structure function $F_2(x, Q^2)$ agree with the OPAL data for the $x$ and $Q^2$ dependence.
Furthermore, our results are also consistent with the predictions obtained from the GRS PDF set for the real photon.
This agreement underscores the predictive power and robustness of our model.
It has been presented that the resulting structure functions for the two cases, $\gamma\gamma$ and $\gamma\rho$, show the similar $x$ and $Q^2$ dependence.
This consistency indicates the natural emergence of VMD in the small $x$ regime, which is derived within our holographic framework rather than imposed as an input. This contrasts with conventional phenomenological approaches, where the vector meson component and its parameters must be introduced explicitly.
However, it is also observed in this study that, in the very small $x$ region, deviations between those two results can clearly be seen.
More experimental data, especially in the smaller $x$ region, are necessary for further constraints on the present descriptions.

Future linear colliders~\cite{Barish:2013yfb} and next-generation linear collider experiments~\cite{CEPCStudyGroup:2023quu}, such as the planned ILC and CEPC, will provide important platforms for deeply exploring the nature of photons. Understanding the intrinsic structure of photons and their role in high-energy scattering processes is one of the most fundamental and challenging topics in high-energy physics. Through these experiments, we can validate the predictions of current theoretical models and further reveal the nature of the strong interaction and the deeper structure of QCD.

\begin{acknowledgments}
This work is supported by the National Natural Science Foundation of China (Grant No. 12475057).
\end{acknowledgments}

\bibliography{ref}

\end{document}